\documentclass[12pt,a4]{article}
\usepackage{amsmath}
\usepackage{amssymb}
\usepackage{amsfonts}
\usepackage{array}
\usepackage{graphicx}
\usepackage{mathrsfs}
\usepackage{multirow}
\usepackage{siunitx}
\usepackage[braket, qm]{qcircuit}
\def\be{\begin{eqnarray}}
\def\ee{\end{eqnarray}}

\begin{document}
\title{Modelling of q-deformed harmonic oscilator on quantum computer}
\author{M. I. Samar and V. M. Tkachuk\\ Professor Ivan Vakarchuk Department for Theoretical Physics, \\ Ivan Franko National University of Lviv,\\ 12 Drahomanov St, Lviv,
UA-79005, Ukraine}
\maketitle

\begin{abstract}
We  present a method of a quantum simulation of a quantum harmonic oscillator in a special case of the deformed commutation relation, which corresponds to the so-called q-deformed oscillator on an IBM quantum computer. Using the method of detection of energy levels of a spin system on a quantum computer by probe spin evolution  we obtain the energy levels of both the q-deformed quantum harmonic and  anharmonic oscillators.
\end{abstract}

\section{Introduction}
Research in string theory and quantum gravity has indicated the presence of a finite lower limit for the achievable resolution of length $\Delta x_{min}$ \cite{GrossMende, Maggiore}. Conversely, on a large scale, the  plane waves or momentum eigenvectors is not well-defined on arbitrary curved spaces. This has led to the proposal that there may also exist a finite lower bound for the achievable resolution of momentum $\Delta p_{min}$ \cite{Kempf:1994}. 
Such nonzero minimal uncertainties can be described in the framework of small corrections to the canonical commutation relation \cite{Kempf1994,KempfManganoMann,HinrichsenKempf,Kempf1997}:
\begin{equation} \label{def}
[x,p] = i\hbar \left(1+ \bar{\alpha}x^2 + \bar{\beta}p^2 \right) \quad 
\end{equation}
with $\bar{\alpha} \neq 0$, $\bar{\beta} \neq 0$, and $\bar{\alpha}\bar{\beta} < \hbar^{-2}$. In such a context, they are given by 
\be
\Delta x_{min} = \frac{\hbar\sqrt{\bar{\beta}}}{\sqrt{(1 - \hbar^2\bar{\alpha}\bar{\beta})}} 
\ee
and 
\be
\Delta p_{min} = \frac{\hbar\sqrt{\bar{\alpha}}}{\sqrt{(1 - \hbar^2\bar{\alpha}\bar{\beta})}} 
\ee
respectively.

Solving problems in quantum mechanics involving the deformed canonical commutation relation can be challenging. In a special scenario where the parameter $\bar{\alpha}$ is zero and $\bar{\beta}$ is greater than zero, the minimum uncertainty in position is given by $\Delta x_{min} = {\hbar}\sqrt{\bar{\beta}}$, while there is no minimum uncertainty in momentum. Consequently, (\ref{def}) can be expressed using momentum space wavefunctions.

Likewise, when $\bar{\alpha}$ is greater than zero and $\bar{\beta}$ is zero, only momentum has a nonzero minimum uncertainty, specifically $\Delta p_{min} = \hbar\sqrt{\bar{\alpha}}$. In this case, Equation (\ref{def}) can be represented using position space wavefunctions.
However, in the general case where both $\bar{\alpha}$ and $\beta$ are greater than zero, there is no representation in either position or momentum space. As a result, one must turn to a Bargmann-Fock space representation  \cite{Kempf:1994}.
 
Despite this, the  exact results for the energy spectrum and the eigenstates of harmonic oscilator  with nonzero minimal uncertainties in both position and momentum was derived using SUSYQM techniques \cite{QT:2003}. However, more complex problems, such as the anharmonic oscillator, have not yet been solved in deformed space with minimal length and minimal momentum. 

The nonlinear oscillator holds a central position in molecular physics, necessitating the consideration of anharmonicities for an accurate description of vibrational dynamics in molecular systems. A commonly used model for characterizing anharmonic spectra involves a one-dimensional harmonic oscillator perturbed by a term containing a quartic power in the coordinate. Extensive research has been dedicated to studying this quantum oscillator with a quartic perturbation. Determining precise energy levels in a quantum mechanical quartic nonlinear oscillator has become a benchmark for testing new methods in quantum mechanics (see for instance, \cite{Matamala}).  

The advancement of quantum programming has led to the widespread usage of techniques for determining the energy levels of quantum systems on quantum computers, including quantum phase estimation algorithm \cite{Kitaev:1995, Abrams:1997, Abrams:1999, Kitaev:1997}, quantum eigenvalue estimation via time series analysis \cite{Somma}, quantum imaginary time evolution algorithms \cite{Motta}, etc.
Recently, new method of detection of energy levels of a spin system on a quantum computer  based on studies of evolution of only one probe spin was proposed \cite{Gnatenko}. The method is stated to be efficient even in the case of noisy quantum devices.

 The paper is organized as follows. In Sect. 2 we consider the harmonic oscillator in context of a special case of the deformed commutation relation (\ref{def}), which corresponds to the so-called q-deformed oscillator. To make it possible to model the q-deformed harmonic oscillator and q-deformed harmonic oscillator  with quadric perturbation on quantum computer we consider the trunctated matrix approach in Sect. 3.  In Section 4 the method for detecting of the energy levels  on the basis of
studies of evolution of probe spin is applied to the  quantum system under consideration and the results of quantum calculations of q-deformed harmonic and anharmonic  oscilator is  on ibmq-pearth are
presented. Conclusions are done in Sect. 5.

\section{q-Deformed harmonic oscillator}
Let us consider the deformed commutation relation (\ref{def}) leading to both minimal length and minimal momentum in the context of harmonic oscillator problem, given by the Hamiltonian 
\be 
H_0=\frac{p^2}{2m}+\frac{m\omega^2x^2}{2}.
\ee
In the dimensionless form the Hamiltonian can be rewritten as
\be H_0=\frac{\hbar\omega}{2}({P^2+X^2  }),\ee
with
\be
X=\frac{x}{a}, \ \  P=\frac{pa}{\hbar},
\ee
and 
$a=\sqrt{\frac{\hbar}{m\omega}} $ denoting the length parameter.

Corresponding  dimensionless deformed commutation relation (\ref{def}) can be written as
\be
[X,P] = i(1 + \alpha X^2 + \beta P^2),
\ee
with  $\alpha=\frac{\tilde\alpha\hbar}{m\omega}$ and  $\beta={\tilde\beta\hbar m\omega}$.

In a special case $\alpha=\beta\neq0$ our problem corresponds to the so-called  q-deformed oscillator\cite{Arik, Quesne}. 
Namely, we can present the position and momentum operators $X$ and $P$ as 
\be
\hat{X}=\frac{\sqrt{1+q}}{2}(\hat{b}^++\hat{b}) \\
\hat{P}=\frac{i\sqrt{1+q}}{2}(\hat{b}^+-\hat{b}),
\ee
where
\be
\hat{b}|{n}\rangle_q=\sqrt{[n]_q}|{n-1}\rangle_q \\
\hat{b}^+|{n}\rangle_q=\sqrt{[n+1]_q}|{n+1}\rangle_q,
\ee
are anihillation and  creation operators
satisfying q-deformed condition 
\be b{b}^+-q{b}^+b = 1,\ee
with 
\be[n]_q=\frac{q^n-1}{q-1} \label{q_number}\ee
 and $q=\frac{1+\alpha}{1-\alpha}$.

 The Hamiltonian of the system and the corresponding energy spectrum  are
\be H =\frac{\hbar\omega}{4}(q + 1)\{b, b^+\},\ee
\be e_n(q)  =E_n/(\hbar\omega)=\frac{1}{4} (q + 1) \left([n]_q + [n + 1]_q\right). \label{e}\ee
Thus  in the $\alpha=\beta\neq0$ case, the harmonic oscillator Hamiltonian with
nonzero minimal uncertainties in position and momentum can be reduced to the q-deformed
oscillator Hamiltonian. 
\section{A truncated matrix approach}
For a harmonic oscillator, the Hamiltonian can be represented as a infinite matrix. However, in the case of low energies, one can truncate the matrix, considering only the ground state and some excited states. This reduces the size of the matrix and simplifies the computations. By taking into account only a limited number of states, you can obtain a reasonably accurate description of the system, including the ground and a few low-lying excited states, while ignoring less significant contributions from higher-energy states.

This approach is particularly useful in quantum mechanics for approximating the energy structure of a system, especially in situations where you are working with low-energy systems that can be effectively described by a limited number of states.

On the other hand, if the truncated matrix has a size of $2^n$, such a system can can be effectively represented using $n$ qubits. The ability to model a system using qubits opens up possibilities for quantum simulations and computations of such a system.

Let us represent our system in $4\times4$ matrix approach.
The creation and anihillation operators has the form
\be
\hat{b}= \begin{pmatrix}
    0  & \sqrt{[1]_q} & 0 &  0 \\
    0  & 0 & \sqrt{[2]_q} &  0 \\
    0  & 0 & 0 & \sqrt{[3]_q} \\
    0  & 0 & 0 &  0 
\end{pmatrix},
\ \
\hat{b}^+= \begin{pmatrix}
    0  & 0 & 0 &  0 \\
    \sqrt{[1]_q}  & 0 & 0 &  0 \\
    0  & \sqrt{[2]_q} & 0 &  0\\
    0  & 0 & \sqrt{[3]_q} & 0  
\end{pmatrix}
\ee
The square of position and momentum operators in considered approach correspondingly are
\be\label{X^2}
X^2=\frac{{[2]_q}}{4} \begin{pmatrix}
    [1]_q  & 0 & \sqrt{[2]_q} &  0 \\
    0  & [1]_q+[2]_q &0 &  \sqrt{[2]_q[3]_q} \\
    \sqrt{[2]_q}  & 0 & [2]_q+[3]_q &  0\\
    0  & \sqrt{[2]_q[3]_q} & 0 & [3]_q+[4]_q 
\end{pmatrix},
\ee
\be
P^2=\frac{{[2]_q}}{4} \begin{pmatrix}
    [1]_q  & 0 & -\sqrt{[2]_q} &  0 \\
    0  & [1]_q+[2]_q &0 &  -\sqrt{[2]_q[3]_q} \\
    -\sqrt{[2]_q}  & 0 & [2]_q+[3]_q &  0\\
    0  & -\sqrt{[2]_q[3]_q} & 0 & [3]_q+[4]_q 
\end{pmatrix}.
\ee
The Hamiltonian in truncated matrix approach can be written as
\be\label{H_0_matrix}
H_0=\frac{\hbar\omega [2]_q}{4} \begin{pmatrix}
    [1]_q  & 0 & 0 &  0 \\
    0  & [1]_q+[2]_q &0 &  0 \\
    0  & 0 & [2]_q+[3]_q &  0\\
    0  & 0 & 0 & [3]_q+[4]_q 
\end{pmatrix}.
\ee
Note, that according to (\ref{q_number}) q-numbers are
\be
[1]_q=1, \\
\ [2]_q=1+q, \nonumber \\ 
\ [3]_q=1+q+q^2, \\
\ [4]_q=1+q+q^2+q^3.
\ee

Now we see that the Hamiltonian of q-deformed harmonic oscillator (\ref{H_0_matrix}) corresponds  to Ising model in magnetic field
\be
H_0= c_1+c_2\sigma^z_2+c_3\sigma^z_1+c_4\sigma^z_1\sigma^z_2,
\ee
with
\be c_1&=&\frac{\hbar\omega [2]_q(2+2[2]_q+2[3]_q+[4]_q)}{16}, \nonumber\\ c_2&=&-\frac{\hbar\omega [2]_q [4]_q}{16},\nonumber\\ c_3&=&-\frac{\hbar\omega [2]_q^4}{16},\nonumber \\ c_4&=&\frac{\hbar\omega [2]_q([4]_q-2[2]_q)}{16}.\label{c}\ee 

In addition we can consider the Hamiltonian 
\be
H_{ho}=H_0+\frac{\gamma}{2} X^2,
\ee
which still describes the same q-deformed harmonic oscillator, 
with the more complicated corresponding spin model, however. 
We obtain the spin Hamiltonian describing  Heisenberg model with antisymmetric exchange
\be \label{spin}
H=d_1+d_2\sigma^z_2+d_3\sigma^z_1+d_4\sigma^z_1\sigma^z_2+d_5\sigma^x_1+d_6\sigma^x_1\sigma^z_2,
\ee
with 
\be
\Tilde{\gamma}=\frac{\gamma}{2m\omega^2}
\ee
\be 
d_1=(1+\Tilde{\gamma})c_1, \\
d_2=(1+\Tilde{\gamma})c_2, \\
d_3=(1+\Tilde{\gamma})c_3, \\
d_4=(1+\Tilde{\gamma})c_4, \\
d_5=\frac{\hbar\omega\Tilde{\gamma}}{8} [2]_q^{\frac{3}{2}}(1+\sqrt{[3]_q}),\\
d_6=\frac{\hbar\omega\Tilde{\gamma}}{8} [2]_q^{\frac{3}{2}}(1-\sqrt{[3]_q}),
\ee

In is interesting that  the Hamiltonian
\be
H_{ao}=H_0+{\delta} X^4,
\ee
describing the quartic anharmonic oscilator
corresponds to the same spin model (\ref{spin}), but with the different coefficient $d_i, i=1,\dots,6.$, given in Appendix A.

\section{Detection of energy levels  on a quantum computer }
Let us study the eigenvalue problem for the Hamiltonian \(H\):
\be
H|\psi\rangle = E|\psi\rangle
\ee

We use the method proposed in \cite{Gnatenko}.
Note that for considered Hamiltonians energy levels are positive ($E>0$).
 We introduce an additional spin (ancilla qubit) and construct the total Hamiltonian in the following form:
\be
H_T = \sigma_0^z H 
\ee
Note that \([\sigma_z^0, H] = 0\), which means that $\sigma_z^0$ and $H$ has common set of eigenstates. Since the eigenvalues of $\sigma_0^z$ are $\pm 1$, the energy spectrum $E_T$ of the total Hamiltonian $H_T$ is  $\pm E$, and is symmetric with respect to $E_T=0$, thus.
It is easy to see that operator $\sigma_x^0$  do anticommute with the total Hamiltonian
$\{\sigma_x^0, H_T\} = 0.$

From this fact we can show that the evolution of the mean value of $\sigma_x^0$ governed by $H_T$ from the initial state 
$|\psi_0\rangle = | + + \dots+\rangle$ reads
\be \langle\sigma_x^0(t) \rangle = \langle + + + \ldots + | \sigma_{x0} e^{-i2H_T t/\hbar} | + + + \ldots + \rangle=\frac{1}{2^{N+1}} \sum_{k=1}^{2^{N+1}} e^{-i2\omega^k_Tt}.\ee
Here  $\omega^k_T$ are the energy levels of total Hamiltonian  in the units of $\hbar$.The frequencies $\omega^k_T$
can be extracted using  the Fourier transformation
\be
\sigma_x^0(\omega) = \frac{1}{2\pi} \int_{-\infty}^{\infty} dt \sigma_x^0(t) e^{i\omega t} = \frac{1}{2^{N+1}} \sum_{k}^{2^{N+1}} \delta\left(\omega - 2\omega^k_{T}\right).
\ee

Quantum protocol for the studies of the evolution of the mean value of a probe spin for $H_{ho}$ or $H_{ao}$ model is presented in Fig. 1. The expressions for angles $\varphi_{1,2},  \lambda_\pm$, $\phi_\pm$, $\theta_\pm$, $\chi_\pm$ are presented in the Appendix A.

By the action of Hadamard gate on each qubit we prepare the initial state $|\psi_0\rangle = | + + \dots+\rangle$. Gate $RY\left(-\frac{\pi}{2}\right)$ in the end of the protocol rotates the state of ancilla qubit to make the meassurement in standard basis. The inner part of the protocol corresponds to the unitary gate $ \exp\left(-i\frac{t \sigma_z^0 H}{\hbar}\right)$.
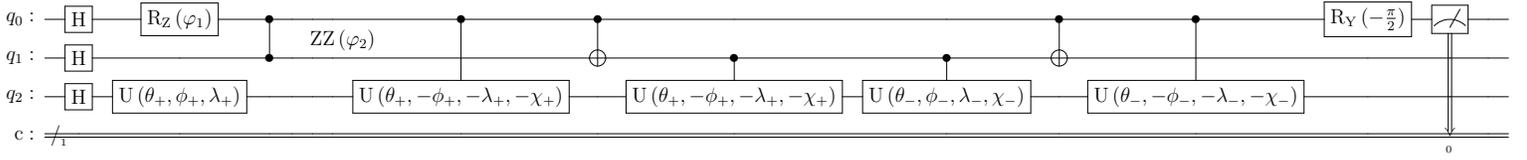
\begin{figure}
\hspace*{-3cm}\scalebox{0.65}{
\Qcircuit @C=1.0em @R=0.2em @!R { \\
	 	\nghost{{q}_{0} :  } & \lstick{{q}_{0} :  } & \gate{\mathrm{H}} & \gate{\mathrm{R_Z}\,(\varphi_1)} & \ctrl{1} & \dstick{\hspace{5.0em}\mathrm{ZZ}\,(\varphi_2)} \qw & \qw & \qw & \ctrl{2} & \ctrl{1} & \qw & \qw & \ctrl{1} & \ctrl{2} & \gate{\mathrm{R_Y}\,(\mathrm{-\frac{\pi}{2}})} & \meter & \qw & \qw\\
	 	\nghost{{q}_{1} :  } & \lstick{{q}_{1} :  } & \gate{\mathrm{H}} & \qw & \control \qw & \qw & \qw & \qw & \qw & \targ & \ctrl{1} & \ctrl{1} & \targ & \qw & \qw & \qw & \qw & \qw\\
	 	\nghost{{q}_{2} :  } & \lstick{{q}_{2} :  } & \gate{\mathrm{H}} & \gate{\mathrm{U}\,(\theta_+,\phi_+,\lambda_+)} & \qw & \qw & \qw & \qw & \gate{\mathrm{U}\,(\theta_+, -\phi_+, -\lambda_+, -\chi_+)} & \qw & \gate{\mathrm{U}\,(\theta_+, -\phi_+, -\lambda_+, -\chi_+)} & \gate{\mathrm{U}\,(\theta_-, \phi_-, \lambda_-, \chi_-)} & \qw & \gate{\mathrm{U}\,(\theta_-, -\phi_-, -\lambda_-, -\chi_-)} & \qw & \qw & \qw & \qw\\
	 	\nghost{\mathrm{{c} :  }} & \lstick{\mathrm{{c} :  }} & \lstick{/_{_{1}}} \cw & \cw & \cw & \cw & \cw & \cw & \cw & \cw & \cw & \cw & \cw & \cw & \cw & \dstick{_{_{\hspace{0.0em}0}}} \cw \ar @{<=} [-3,0] & \cw & \cw\\ 
\\ }}
\caption{\footnotesize{Quantum protocol for studies of the evolution of the mean value of $\sigma_x^0$ for $H_{ho}$ or $H_{ao}$ model}}
\end{figure}

Quantum protocol Fig. 1 was implemented on ibmq-quito.
Since, the value $\sigma_x^0(\omega)$ is real,  in Fig. 2, we present the real part of $\sigma_x^0(\omega)$ detected on the quantum device ibmq-quito. The sharp peaks of the $\Re\sigma_x^0(\omega)$ presented in Fig. 2 corresponds to the positive energy levels of total Hamiltonian $H_T$.

		\begin{figure}[h!]
			\centering
			\includegraphics[width=10cm]{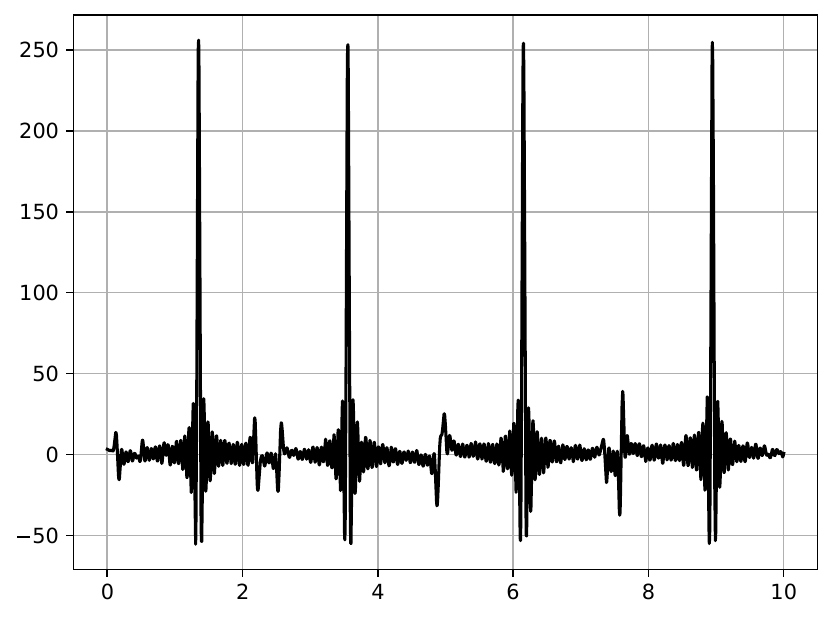}
			\caption{\footnotesize{$\Re\sigma_x^0(\omega)$ detected on the \text{ibmq-quito.}}}
			\label{fig.1}
		\end{figure}
 The results of  the detection of the energy levels of the q-deformed harmonic oscillator depending on $q$ for different values of parameter $\gamma$ are presented on Fig. 3. Analytical values of energy spectrum for $H_{ho}$ can be found from (\ref{e}) with $\omega \rightarrow \sqrt{\omega^2+\gamma}$. From Fig. 3 we see that analitical results (solid lines) are in good agreement with the results of quntum computations (dots). This fact is an additional confirmation of the efficiency of our method.

 We also have detected the first four energy levels for quadric anharmonic q-deformed oscillator. Dependence of first four energy levels on $q$ for harmonic oscilator $H_{ao}=H_0+{\delta} X^4$ is presented in Figures 4 and 5.

	\begin{figure}[h!]
		\centering
		\includegraphics[width=10cm]{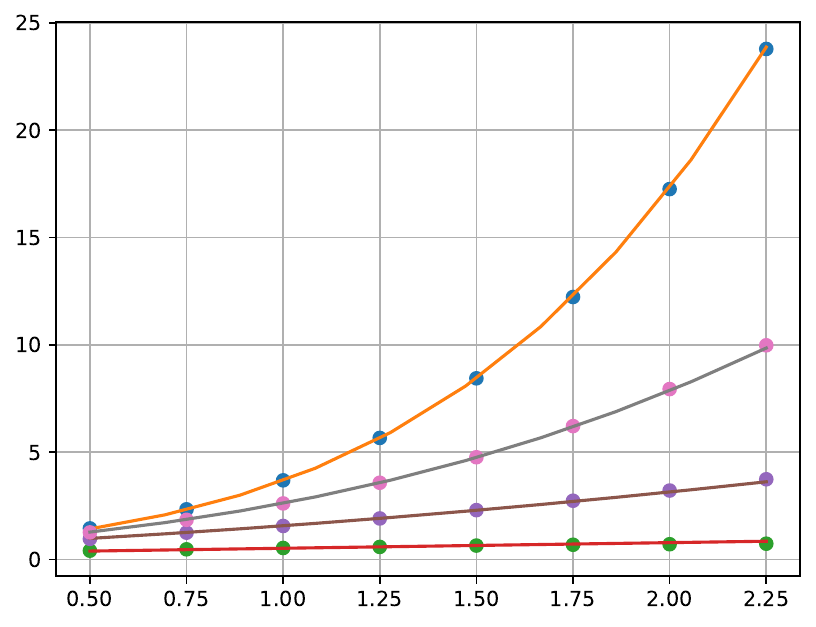}
  \includegraphics[width=10cm]{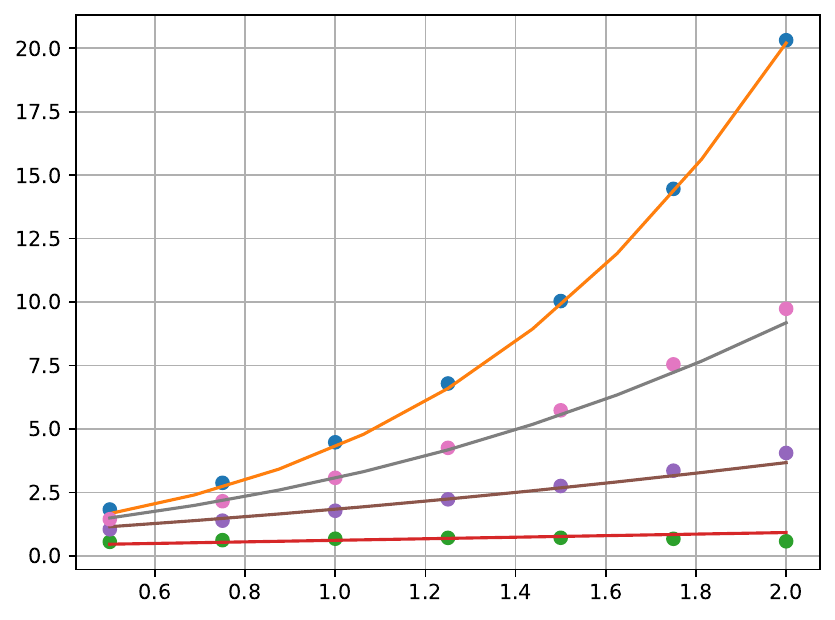}
  \includegraphics[width=10cm]{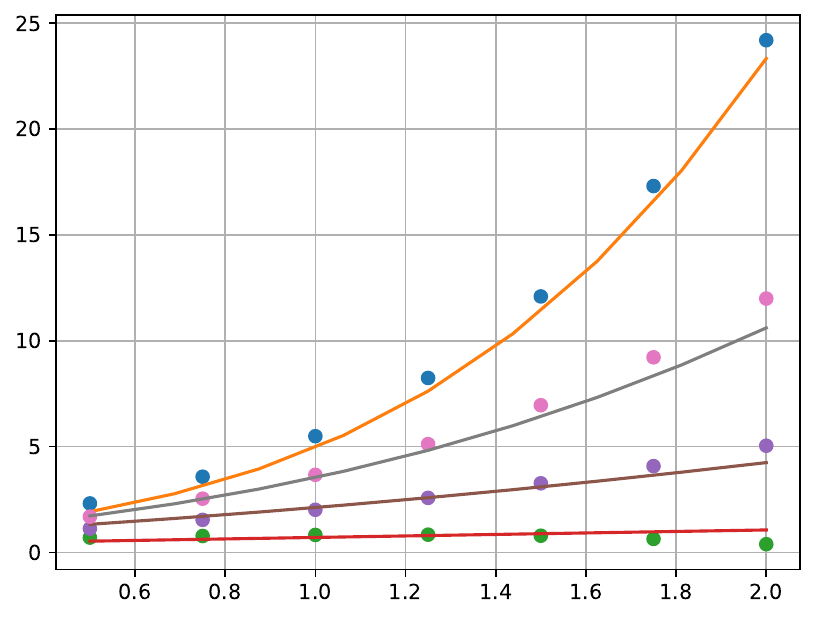}
		\caption{\footnotesize{Dependences of first four energy levels on $q$ for harmonic oscilator $H=H_0+\frac{\gamma}{2} X^2$, $\gamma=0.1; 0.5; 1$. The solid lines present the analytical results, dots presents calculations on a quantum computer \text{ibmq-quito.}}}
		\label{fig.2}
	\end{figure}
\newpage

	\begin{figure}[h!]
		\centering
		\includegraphics[width=10cm]{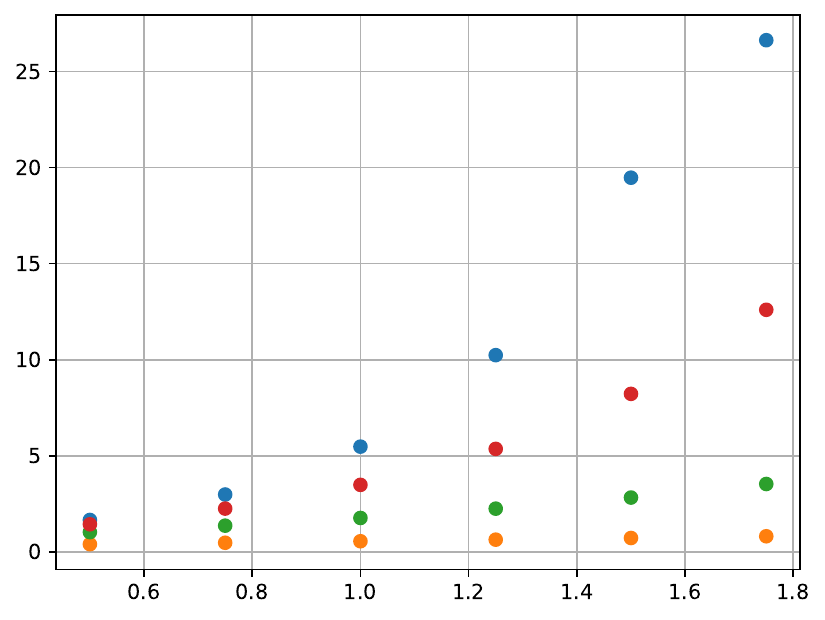}
		\caption{\footnotesize{Dependence of first four energy levels on $q$ for harmonic oscilator $H=H_0+\delta x^4$, $\delta=0.1$. Dots presents calculations on a quantum computer \text{ibmq-quito.}}}
		\label{fig.2}
	\end{figure}

	\begin{figure}[h!]
		\centering
		\includegraphics[width=10cm]{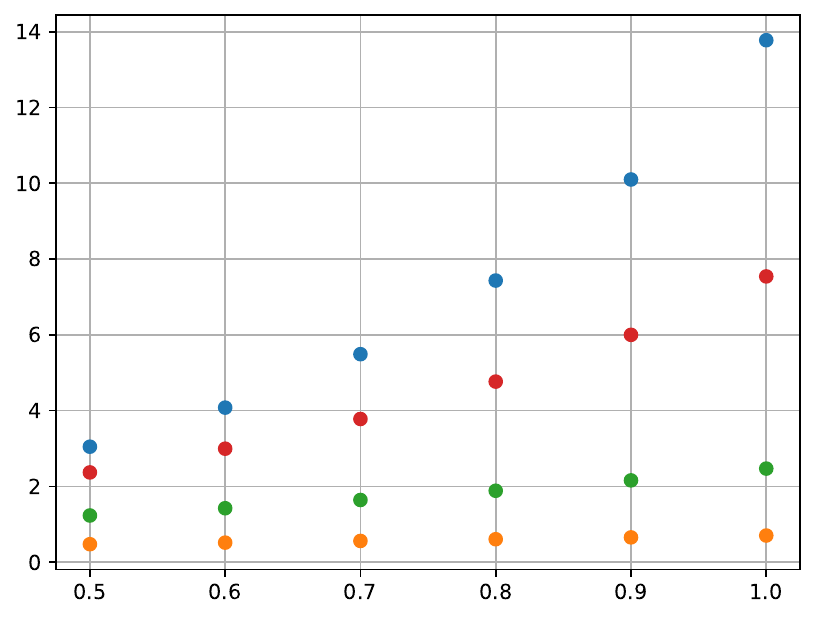}
		\caption{\footnotesize{Dependence of first four energy levels on $q$ for harmonic oscilator $H=H_0+\delta x^4$, $\delta=0.5$. Dots presents calculations on a quantum computer \text{ibmq-quito.}}}
		\label{fig.2}
	\end{figure}

\newpage
\section{Conclusion}
The method for detecting energy levels in a spin system, based on the evolution study of only one probe spin, has been applied to the q-deformed harmonic oscillator in a truncated matrix approach. The proposed algorithm was implemented on IBM's quantum computer, ibmq-quito, successfully detecting the first four energy levels of the q-deformed harmonic oscillator, which showed good agreement with analytical results. Additionally, we have studied the quadric anharmonic q-deformed oscillator. Surprisingly, the spin model corresponding to this system is the same as for the q-deformed harmonic oscillator. Thus, using the same algorithm on IBM's quantum computer ibmq-quito, we obtained, for the first time, the first four energy levels for the quadric anharmonic q-deformed oscillator.

\section{Acknowledgement}
This work was supported by the Project 2020.02/0196 (No. 0120U104801) from National Research Foundation of Ukraine.

\appendix

\newpage
\appendix
\section{Appendix}
\numberwithin{equation}{section}
\setcounter{equation}{0}
Let us calculate the $X^4$ operator in the truncated matrix approach up to $4\times4$ dimension of the matrix. We obtain 
\be\label{X^4}
{ X^4=\frac{{[2]_q^2}}{16} \begin{pmatrix}
    1+[2]_q  & 0 & \sqrt{[2]}(1+[2]_q+[3]_q) &  0 \\
    0  & \begin{aligned} (1+[2]_q)^2+\\[-0.8ex]{}+[2]_q[3]_q \end{aligned}&0 &  \begin{aligned} \sqrt{[2]_q[3]_q}(1+\\[-0.8ex]{}+[2]_q+[3]_q+[4]_q)\end{aligned} \\
    \begin{aligned} \sqrt{[2]}(1+[2]_q+[3]_q)  \end{aligned}& 0 &\begin{aligned} ([2]_q+[3]_q)^2+\\[-0.8ex]{}+[2]_q+[3]_q[4]_q  \end{aligned} &  0\\
    0  &\begin{aligned} \sqrt{[2]_q[3]_q}(1+\\[-0.8ex]{}+[2]_q+[3]_q+[4]_q)\end{aligned} & 0 & \begin{aligned} ([3]_q+[4]_q)^2+\\[-0.8ex]{}+[2]_q[3]_q+[4]_q[5]_q \end{aligned} & 
\end{pmatrix}}
\ee
Comparing the matrix for $X^2$ given in (\ref{X^2}) with the matrix for $X^4$ given in
 (\ref{X^4}) we see that this matrices have the same structure. This fact leads to the conclusion that spin models corresponding to the q-deformed harmonic oscillator and quadric anharmonic oscillator are the same.
 
 Denoting the elements of matrix $X^4$ as $A_{ij}$, we introduce 
 \be c'_1=\frac{1}{4}(A_{11}+A_{22}+A_{33}+A_{44}), \\
  c'_2=\frac{\delta}{4}(A_{11}-A_{22}+A_{33}-A_{44}),\\
  c'_3=\frac{\delta}{4}(A_{11}+A_{22}-A_{33}-A_{44}), \\
  c'_4=\frac{\delta}{4}(A_{11}-A_{22}-A_{33}+A_{44}), \\
  c'_5=\frac{\delta}{2}(A_{13}+A_{24}),\\
  c'_6=\frac{\delta}{2}(A_{13}-A_{24}).
 \ee
The parameters of the spin Hamiltonian  $d_i$ thus can be written as 
\be d_i=c_i+c'_i, \ee
 with $c_1, c_2, c_3$ and $c_4$ are given in (\ref{c}) and $c_5=c_6=0$.

The angles which appears in the quantum protocol given in Fig. 1 can be expressed as
\be \varphi_1=2d_1(q)t/\hbar, \  \varphi_2=2d_2(q)t/\hbar \ee
\be \lambda_\pm(q,t)= -\arctan(\alpha_\pm(q)\cot(J_\pm(q,t)))\ee
\be \phi_\pm(q,t) = \lambda_\pm(q,t)+\pi,  \ \ \chi_\pm(q,t)= -\lambda_\pm(q,t)+\pi/2\ee
\be \theta_\pm(q,t)= 2\arcsin(\beta_\pm(q)\sin(J_\pm(q,t))),\ee
where we use notations
\be  {j_\pm}(q) = \sqrt{\left({d_3}\left(q\right) \pm {d_4}\left(q\right) \right)^{{2}} + \left({d_5}\left(q\right) \pm d_6\left(q\right) \right)^{{2}}},\ee
\be \alpha_\pm(q) = \frac{\mathrm{d_3}\left(q\right)\pm \mathrm{d_4}\left(q\right)}{\mathrm{j_\pm}\left(q\right)}, \ \ \beta_\pm(q) = \frac{\mathrm{d_5}\left(q\right) \pm \mathrm{d_6}\left(q\right)}{\mathrm{j_\pm}\left(q\right)},  \ee
\be  J_\pm(q,t)= \frac{j_\pm(q)t}{\hbar}. \ee


\begin{thebibliography}{99}

\bibitem{GrossMende}D. J. Gross  and P. F. Mende,  Nucl. Phys. B 303, 407 (1988).
\bibitem{Maggiore}M. Maggiore,  Phys. Lett. B 304, 65 (1993).
\bibitem{Kempf:1994} Kempf, A. Czech J Phys 44, 1041–1048 (1994). 
\bibitem{Kempf1994}A. Kempf, J. Math. Phys. 35, 4483 ( 1994).
\bibitem{KempfManganoMann} A. Kempf, G. Mangano  and R. B. Mann, Phys. Rev. D 52, 1108 (1995).
\bibitem{HinrichsenKempf}H. Hinrichsen  and A. Kempf, J. Math. Phys. 37, 2121 (1996).
\bibitem{Kempf1997} A. Kempf, J. Phys. A 30, 2093 (1997).
\bibitem{QT:2003} C. Quesne  and V. M. Tkachuk, J. Phys. A {\bf 36}, 10373-10391 (2003). 
\bibitem{Matamala} A. R. Matamala, C. R. Maldonado, Phys. Let. A, {\bf 308}  319-322 (2003).
\bibitem{Kitaev:1995} A. Kitaev, , arXiv:quant-ph/9511026 (1995)
\bibitem{Abrams:1997} D.S. Abrams, S. Lloyd,  Phys. Rev. Lett. 79, 2586 (1997)
\bibitem{Abrams:1999} D.S. Abrams, S. Lloyd, . Phys. Rev. Lett. 83, 5162(1999)
\bibitem{Kitaev:1997} A.Y. Kitaev,  Russian Math. Surv. 52, 1191 (1997)
\bibitem{Somma} R.D. Somma,  New J. Phys. 21, 123025 (2019)
\bibitem{Motta} M. Motta, Ch. Sun, A.T.K. Tan et al.,  Nat. Phys. 16, 205 (2020)
\bibitem{Gnatenko} Kh. P. Gnatenko , H. P. Laba, V. M. Tkachuk, Eur. Phys. J. Plus (2022) 137:522
\bibitem{Arik} Arik M and Coon D D 1976 J. Math. Phys. 17 524
\bibitem{Quesne} Quesne C, Penson K A and Tkachuk V M 2003 Phys. Lett. A 313 29

\end{thebibliography}
\end{document}